\documentclass[prb,twocolumn,showpacs,float]{revtex4-1}
\usepackage{graphicx}
\usepackage{amssymb}
\usepackage{amsmath}
\usepackage{xspace}
\usepackage{url}
\usepackage{subfigure}

\begin{document} 
\title{Mie scattering analog in graphene: lensing, \\ particle confinement, and depletion of Klein tunneling}

\author{R. L. Heinisch, F. X. Bronold, and 
H. Fehske}
\affiliation{Institut f{\"ur} Physik,
             Ernst-Moritz-Arndt-Universit{\"a}t Greifswald,
             17487 Greifswald,
             Germany}

\date{\today}
\begin{abstract}
Guided by the analogy to Mie scattering of light on small particles we show that the propagation of a Dirac-electron wave in graphene can be manipulated by a circular gated region acting as a quantum dot. Large dots enable electron lensing, while for smaller dots resonant scattering entails electron confinement in quasibound states. Forward scattering and Klein tunneling can be almost switched off for small dots by a Fano resonance arising from the interference between resonant scattering and the background partition.
 
\end{abstract}
\pacs{}
\maketitle

\section{introduction}

A counterintuitive feature of relativistic particles is the unimpeded penetration through high and wide potential barriers, termed Klein tunneling.\cite{Klein28} Rooted in negative energy states in the barrier, Klein tunneling is caused in graphene by a particularity of the band structure.\cite{KNG06} The intersection of energy bands at the edge of the Brillouin zone leads to a gapless conical energy spectrum in two inequivalent valleys. Low-energy quasiparticles are described by a massless Dirac equation\cite{Dirac28} and have a pseudospin which gives the contribution of the two sublattices of the graphene honeycomb lattice to their make-up. Chirality---the projection of the pseudospin on the direction of motion---is  responsible for the conservation of pseudospin  and the absence of backscattering  for potentials diagonal in sublattice space.\cite{ANS98}
This allows an experimental verification of  Klein tunneling  in a solid state system. Early resistance measurements across a barrier indicated Klein tunneling\cite{HSS07,SHG09,GMS08} and in a conclusive experiment Klein tunneling was revealed in the phase shift of the conductance fringes at low magnetic field.\cite{YK09}

Due to Klein tunneling electrons can usually not be confined electrostatically. Thus, designs for circuitry taken from traditional semiconductor-based electronics cannot be used in graphene and unconventional electronic devices are designed on optical analogs.\cite{CFA07,PSY08,WLL11}  This has lead to the proposal of a potential step as a lens for propagating electron beams \cite{CFA07} or the experimental implementation of the counterpart of an optical fiber cable.\cite{WLL11}

In the present work we study, guided by the analogy to Mie scattering of light on small particles,\cite{Mie08,BW99} the scattering of a plane electron wave on a circular potential step in graphene. 
The set-up we consider is a plane graphene sheet on a gated substrate with a separate circular region to tune the potential step by applying a bias. Circular dots have mostly been analyzed in etched structures with a focus on electron confinement as potential hosts for spin qbits.\cite{WRA09,PSK08,MMP11,RT10} Moreover, single dots and voids\cite{HG07} as well as multiple dots arranged in a corral\cite{VAW11} have been used to model the scattering by impurities or metallic islands placed on a graphene sheet. The configuration we consider has been studied for bound states in unbiased graphene\cite{BTB09,TOG10} and for its ray optical scattering properties in single and bilayer graphene.\cite{CPP07,PPC09} 
We show that the dot operates as an electron switch with the preferred scattering angles controlled by its size and applied bias. For small dots forward scattering and Klein tunneling can be almost completely suppressed due to a Fano resonance\cite{Fano61} between the background partition and the resonant contribution to electron scattering.   

\section{theory}

For a gate potential that is smooth on the scale of the lattice constant but sharp on the scale of the de Broglie wavelength the low-energy electron dynamics is described by the single-valley Dirac-Hamiltonian 
\begin{equation}
H=-i\nabla\mathbf{\sigma}+V\theta (R-r),
\end{equation}
 where R is the radius, \(V\) the applied bias of the gated region, and \(\mathbf{\sigma}=(\sigma_x,\sigma_y)\) are Pauli matrices. We use units such that \(\hbar=1\) and the Fermi  velocity \(v_\mathrm{F}=1\). We do not consider effects of disorder or edge roughness of the step, which are beyond the present continuum approach. Our focus is on the analogy to optics. The difference between Dirac's equation, governing a two-spinor wavefunction in 2D, and Maxwell's equations in 3D limits this analogy. In particular Klein tunneling, which is also found for a smooth gate potential \cite{CF06} or from a tight-binding calculation,\cite{LBR12} has no optical equivalent. Known phenomena from Mie scattering will appear in graphene in new guise to satisfy the absence of backscattering. 
 To solve the scattering problem, we expand the incident plane wave  in eigenfunctions in polar coordinates,
 \begin{equation}
 \psi^\mathrm{in}=\frac{1}{\sqrt{2}}\left(
 \begin{matrix}e^{ikx} \\
\alpha e^{ikx}\end{matrix}\right) = \sum_{m=-\infty}^\infty \sqrt{\pi} i^{m+1} \psi_m^{(1)}(kr)
 \end{equation}
  and match the reflected wave
  \begin{equation}
\psi^\mathrm{ref}=\sum_{m=-\infty}^\infty  \sqrt{\pi} i^{m+1} a_m^\mathrm{r} \psi_m^{(3)}(kr)
  \end{equation}
  and the transmitted wave
  \begin{equation}
  \psi^\mathrm{trans}=\sum_{m=-\infty}^\infty  \sqrt{\pi} i^{m+1} a_m^\mathrm{t} \psi_m^{(1)}(qr)
  \end{equation}
  (\(q\) is the wave vector inside the gated region)  term by term so that continuity of the wavefunction is satisfied.  The eigenfunctions in polar coordinates to an energy \(E=\alpha k\), where the band index \(\alpha=1\) denotes the conduction and \(\alpha=-1\) the valence band, read
 \begin{equation}
 \psi_m^{(1,3)}=\frac{1}{\sqrt{2\pi}}\left(
\begin{matrix} 
-i \mathcal{Z}_m^{(1,3)}(kr) e^{im\phi} \\ \alpha \mathcal{Z}_{m+1}^{(1,3)}(kr)e^{i(m+1)\phi}
\end{matrix}\right)
 \end{equation}
  with \(\mathcal{Z}_m^{(1)}=J_m\) Bessel's or  \(\mathcal{Z}_m^{(3)}=H_m^{(1)}\)  Hankel's function of the first kind.\cite{BM87} The scattering coefficients are given by
\begin{equation}
a_m^\mathrm{r}=-\frac{J_{m+1}(N\rho)J_m(\rho)-\alpha \alpha^\prime J_m(N\rho)J_{m+1}(\rho)}{J_{m+1}(N\rho)H_m^{(1)}(\rho)-\alpha \alpha^\prime J_m(N\rho) H_{m+1}^{(1)}(\rho)}  \text{ ,}
\end{equation} 
and the transmission coefficients by
\begin{equation}
a_m^\mathrm{t}=\frac{H_{m+1}^{(1)}(\rho)J_m(\rho)-H_m^{(1)}(\rho)J_{m+1}(\rho)}{H_{m+1}^{(1)}(\rho)J_m(N\rho)-\alpha \alpha^\prime H_m^{(1)}(\rho) J_{m+1}(N\rho)}  \text{ ,}
\end{equation}
where we introduced the size parameter \(\rho=kR\) and the modulus of the refractive index \(N=|V-E|/|E|\), and \(\alpha^\prime\) is the band index inside the gated region. These coefficients satisfy the relations \(a_{-m}=a_{m-1}\) and \(b_{-m}=\alpha \alpha^\prime b_{m-1}\). The electron density is given by \(n=\psi^\dagger \psi\) and the current by \(\mathbf{j}=\psi^\dagger \sigma \psi\) where \(\psi=\psi^\mathrm{in}+\psi^\mathrm{ref}\) outside and \(\psi=\psi^\mathrm{trans}\) inside the gated region. The far-field radial component of the reflected current, which characterizes angular scattering, reads
\begin{align}
j_r^\mathrm{ref}=\frac{4}{\pi k r}  \sum_{m,m^\prime=0}^\infty  a_{m^\prime}^{\mathrm{r}\ast} a_m^\mathrm{r} [ & \cos ((m^\prime+m+1)\phi)  \nonumber \\  &+\cos((m-m^\prime)\phi) ].
\end{align} 
The scattering efficiency, that is, the scattering cross section divided by the geometric cross section is given by \begin{equation}
Q=\frac{4}{\rho}\sum_{m=0}^\infty |a_m^\mathrm{r}|^2.
\end{equation}

\section{results}

\begin{figure}
{\includegraphics[width=1\linewidth]{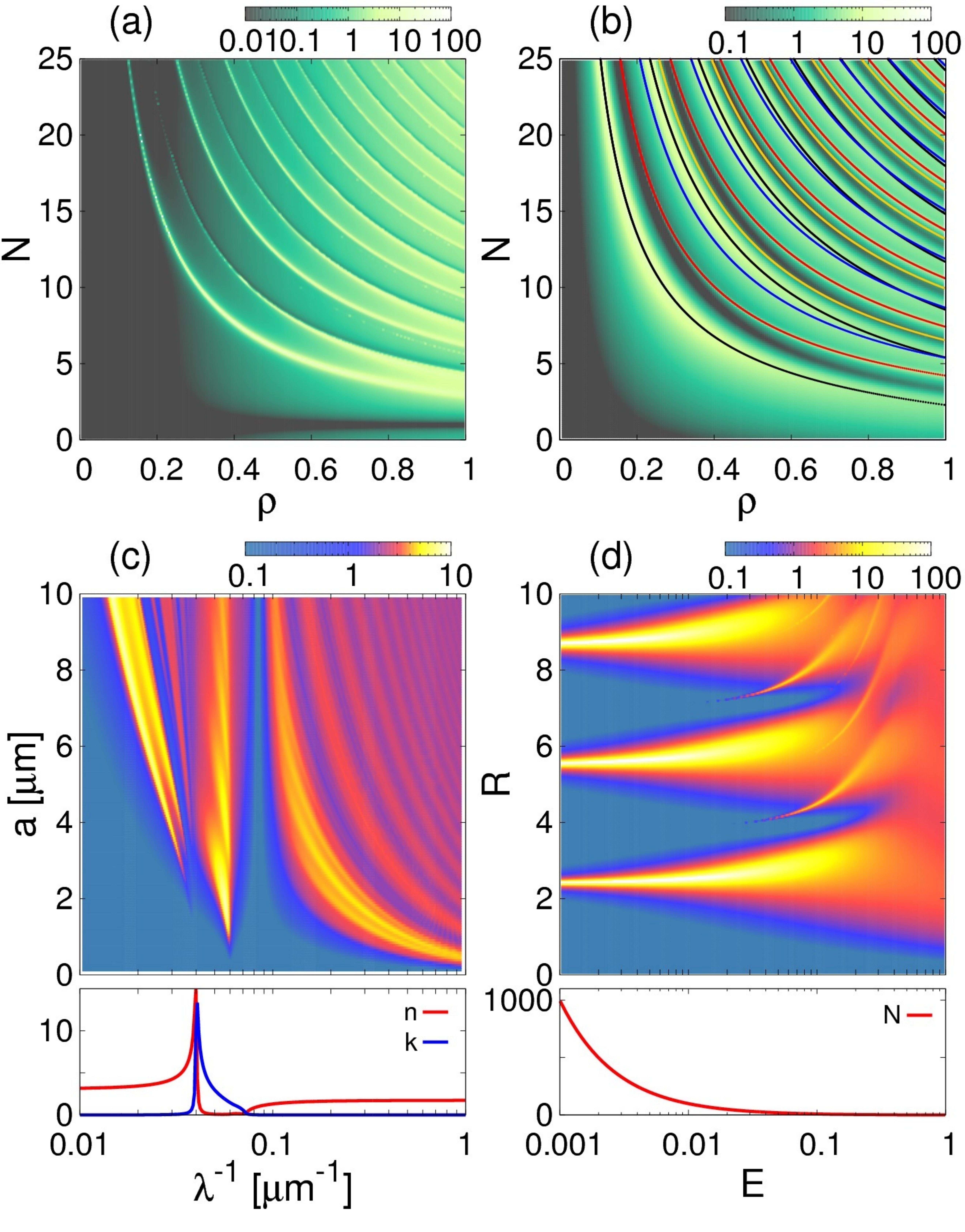}}

\caption{(color online) Comparison of optical Mie scattering and scattering by a gated region in graphene. (a) Scattering efficiency \(Q\) for Mie scattering for positive refractive index and (b) for a gated region in graphene with \(E<V\) as a function of \(N\) and \(\rho\). In (b) the maxima of the scattering coefficients \(a_0\), \(a_1\) \(a_2\), and \(a_3\) are indicated by the black, red, blue and yellow lines. Note that \(Q\neq 0\) in graphene whereas \(Q=0\) in optical Mie scattering for \(N=1\). The reason is that the refractive index in graphene is negative which is already apparent in the analog of Snell's refraction law.\cite{CFA07} (c) \(Q\) for Mie scattering as a function of the particle radius \(a\) and the inverse wavelength \(\lambda^{-1}\) for MgO (complex refractive index \(N=n+ik\) below). (d) \(Q\) in graphene as a function of \(R\) and \(E\) for \(V=1\) (\(N\) for \(V=1\) below).}
\label{figure1}
\end{figure}

 The similar form of the scattering coefficients \(a_m\) in optical Mie scattering\cite{Mie08,BW99} and in graphene should allow for the same scattering phenomena. Indeed, the scattering efficiency as a function of \(\rho\) and \(N\) shows the same trends: \(Q\) features sharp resonances due to the excitation of normal modes for large \(N\) and small \(\rho\) and a broad and overlapping ripple structure for larger \(\rho\) (cf. Fig.~\ref{figure1}a,b).

Which of these behaviors is realized depends on the actual values of the refractive index. In optics the polarizable excitations of the solid determine \(N\). For instance for magnesium oxide (Fig.~\ref{figure1}c), a dielectric with a strong phonon resonance, the real (imaginary) part of  \(N\) is large only below (above) the phonon resonance which entails ordinary (anomalous) optical resonances.\cite{BW99,TL06,HBF12} Over most of the spectrum \(N\) is small which leads to a broad ripple structure. In stark contrast, \(N\) in graphene is tied to the ratio of wavenumbers. A small \(E\) leads to a large \(N\) while  \(E\rightarrow V\) entails \(N \rightarrow 0\). Thus, for low \(E\) very sharp resonances are realized which broaden and overlap for larger energies (Fig.~\ref{figure1}d).

\begin{figure*}
\begin{minipage}{0.32\linewidth}
{\includegraphics[width=1.15\linewidth]{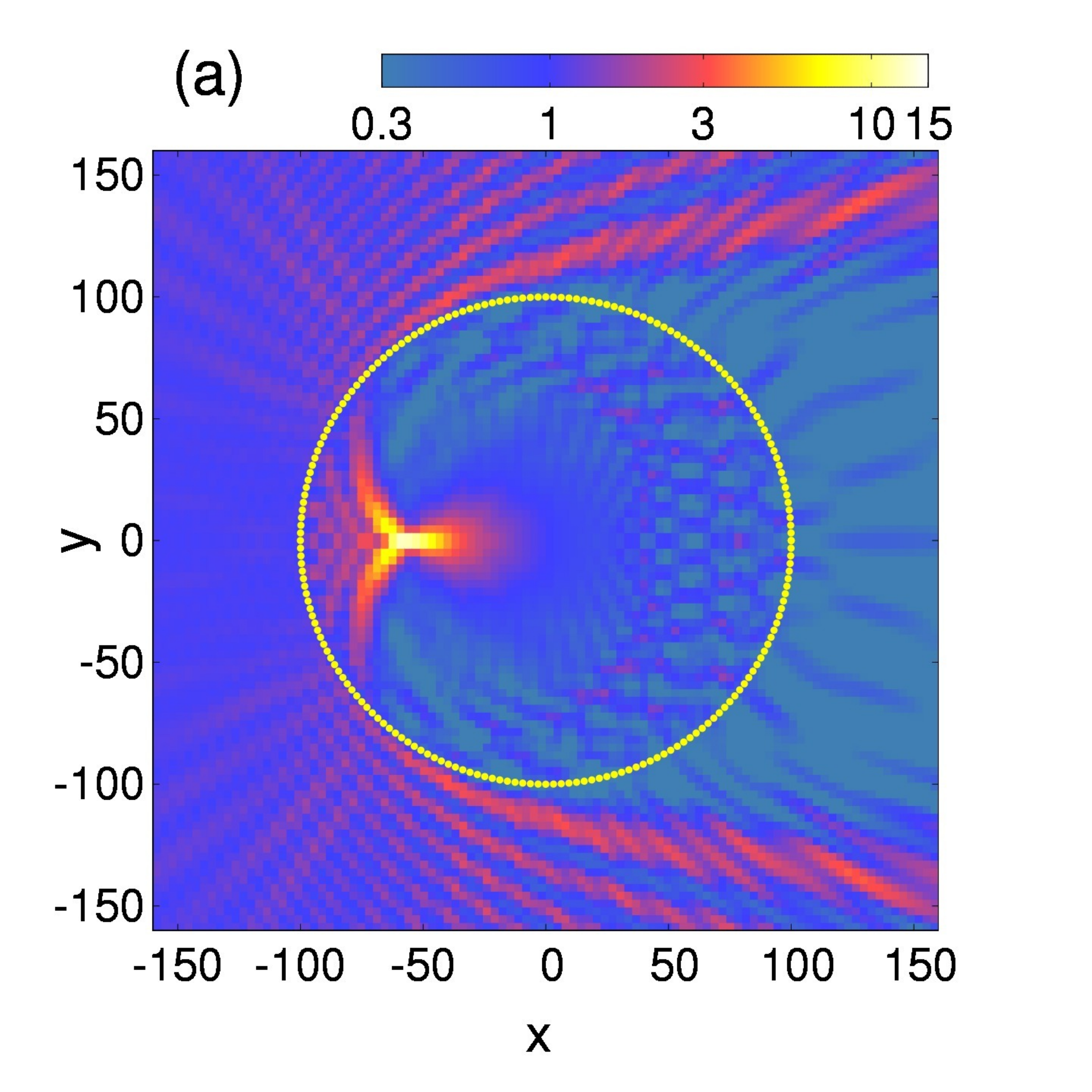}}
\end{minipage}
\begin{minipage}{0.32\linewidth}
{\includegraphics[width=1.15\linewidth]{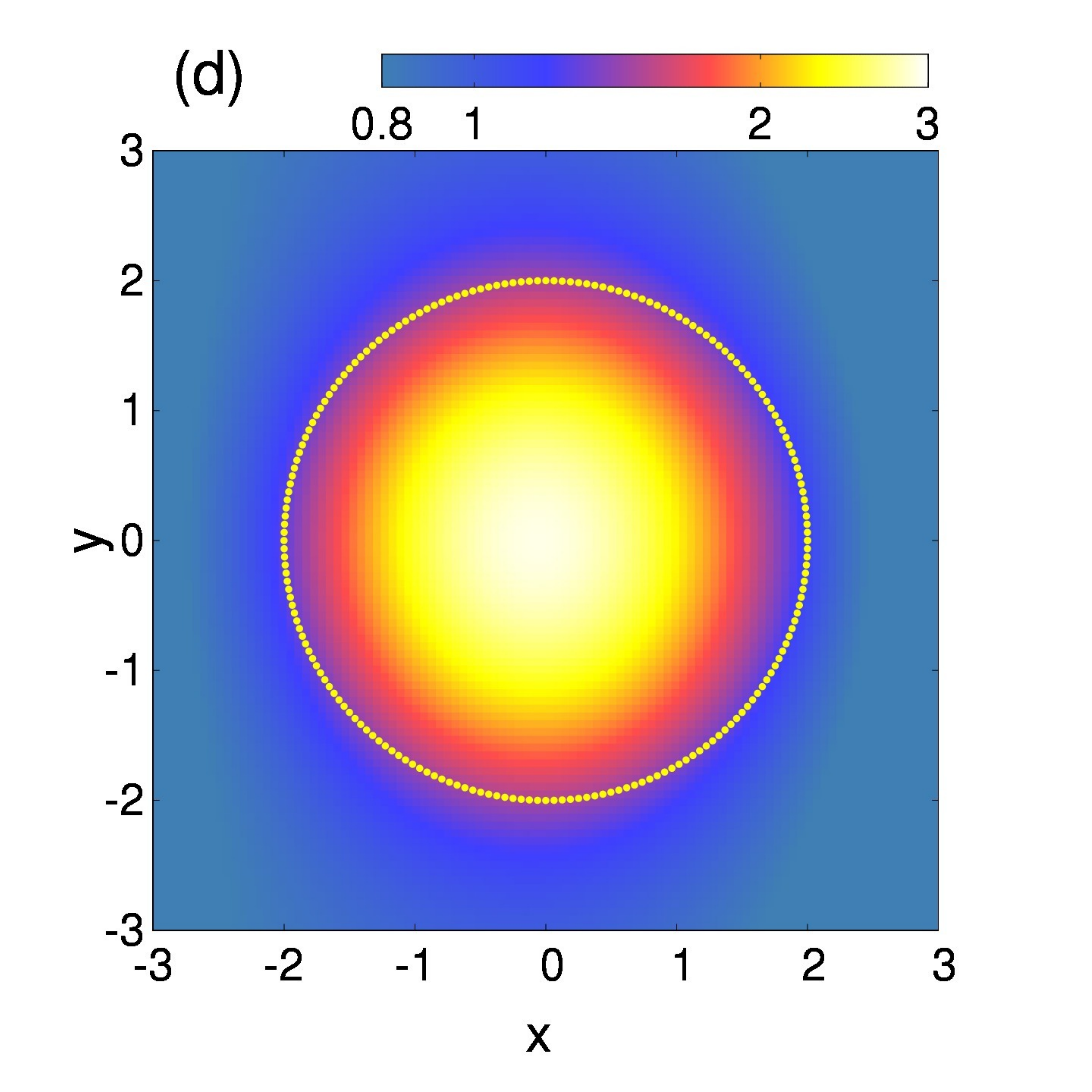}}
\end{minipage}
\begin{minipage}{0.32\linewidth}
{\includegraphics[width=1.15\linewidth]{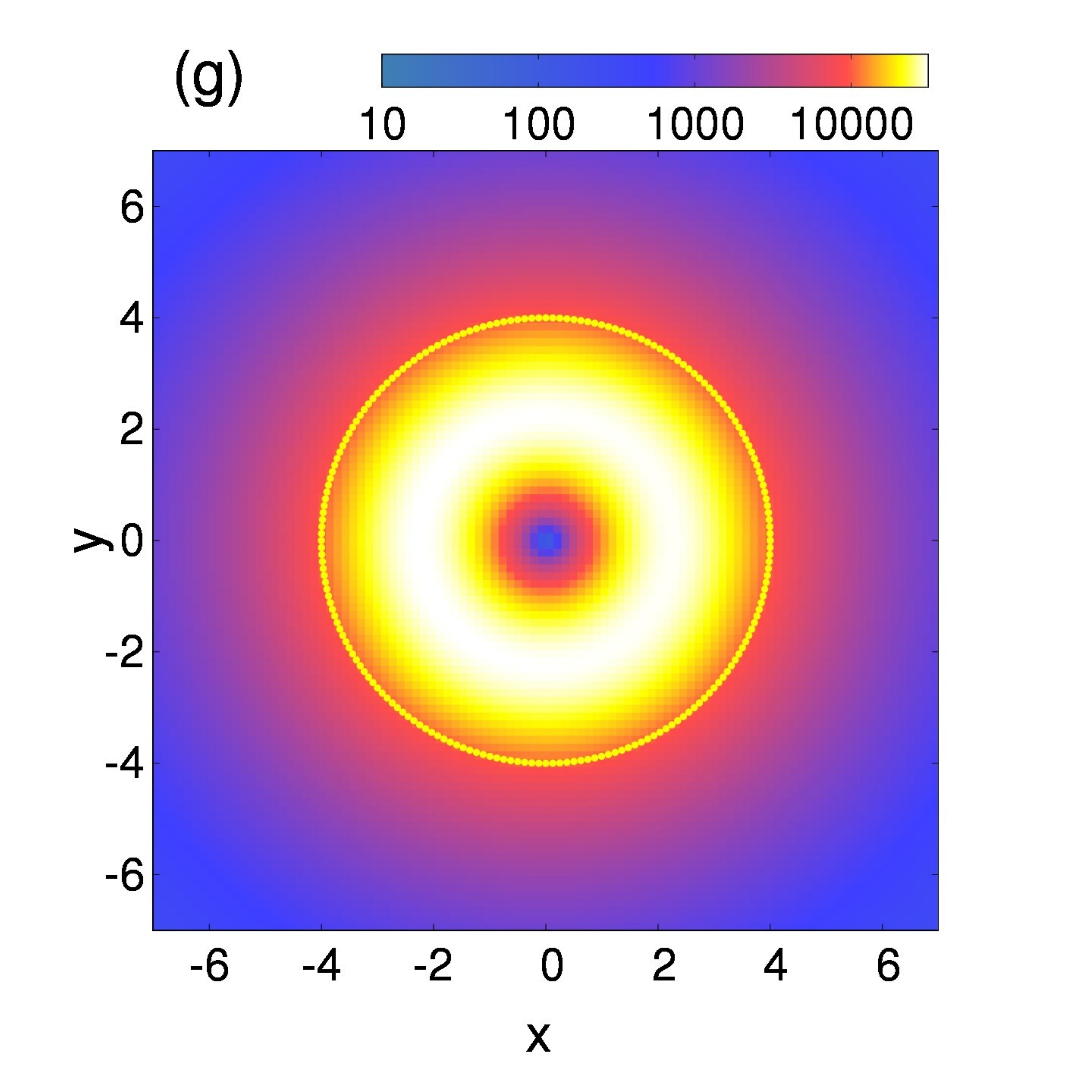}}
\end{minipage}

\begin{minipage}{0.32\linewidth}
{\includegraphics[width=1.15\linewidth]{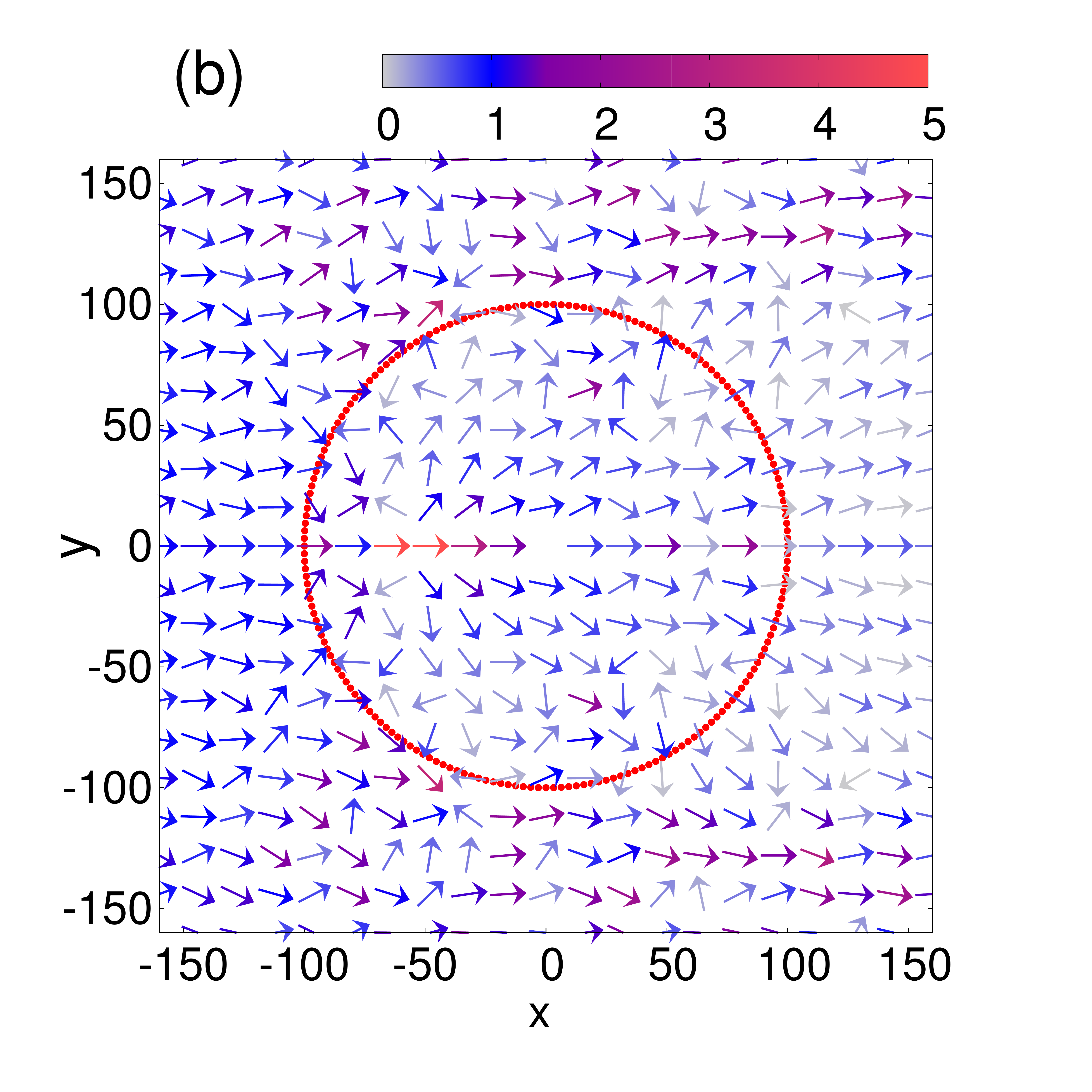}}
\end{minipage}
\begin{minipage}{0.32\linewidth}
{\includegraphics[width=1.15\linewidth]{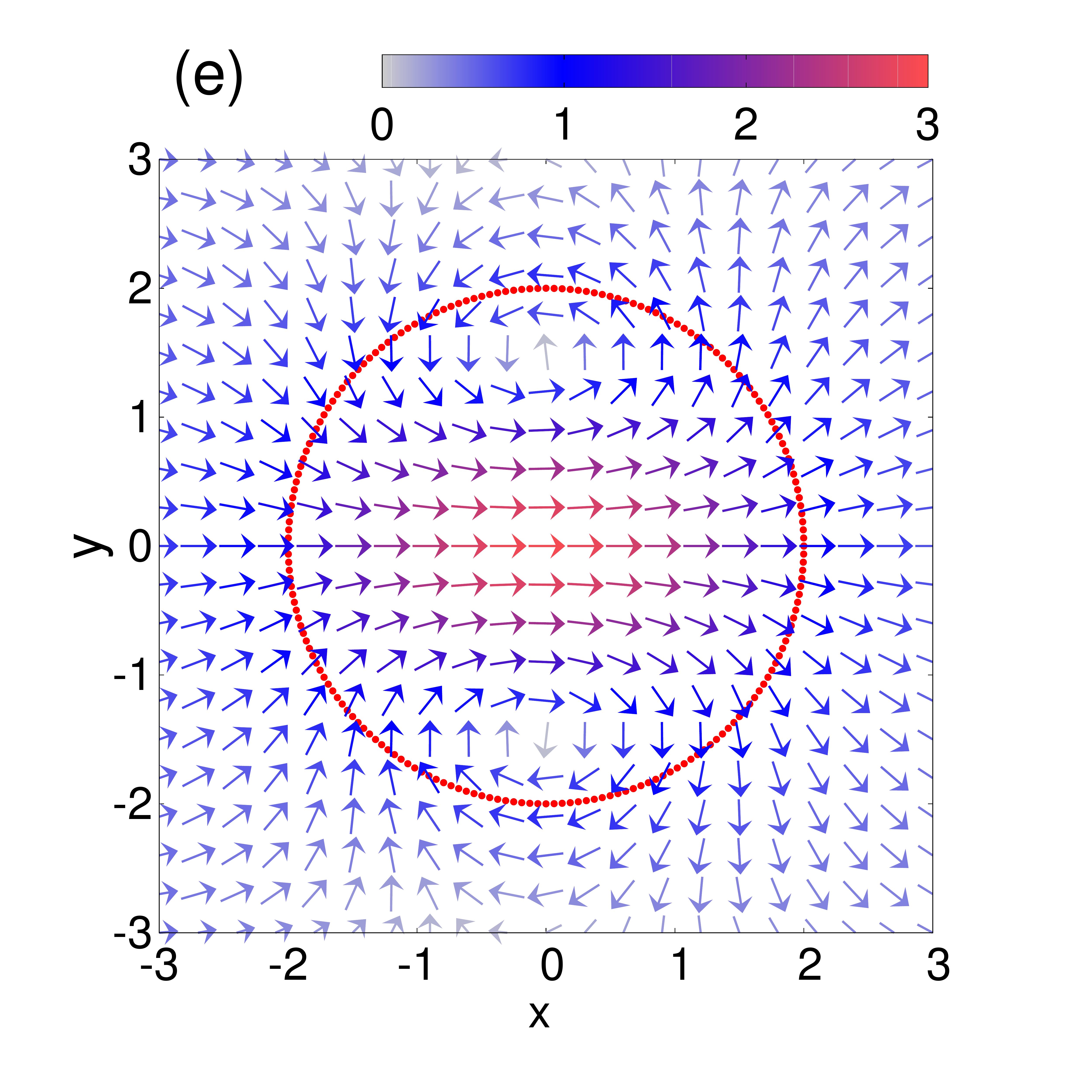}}
\end{minipage}
\begin{minipage}{0.32\linewidth}
{\includegraphics[width=1.15\linewidth]{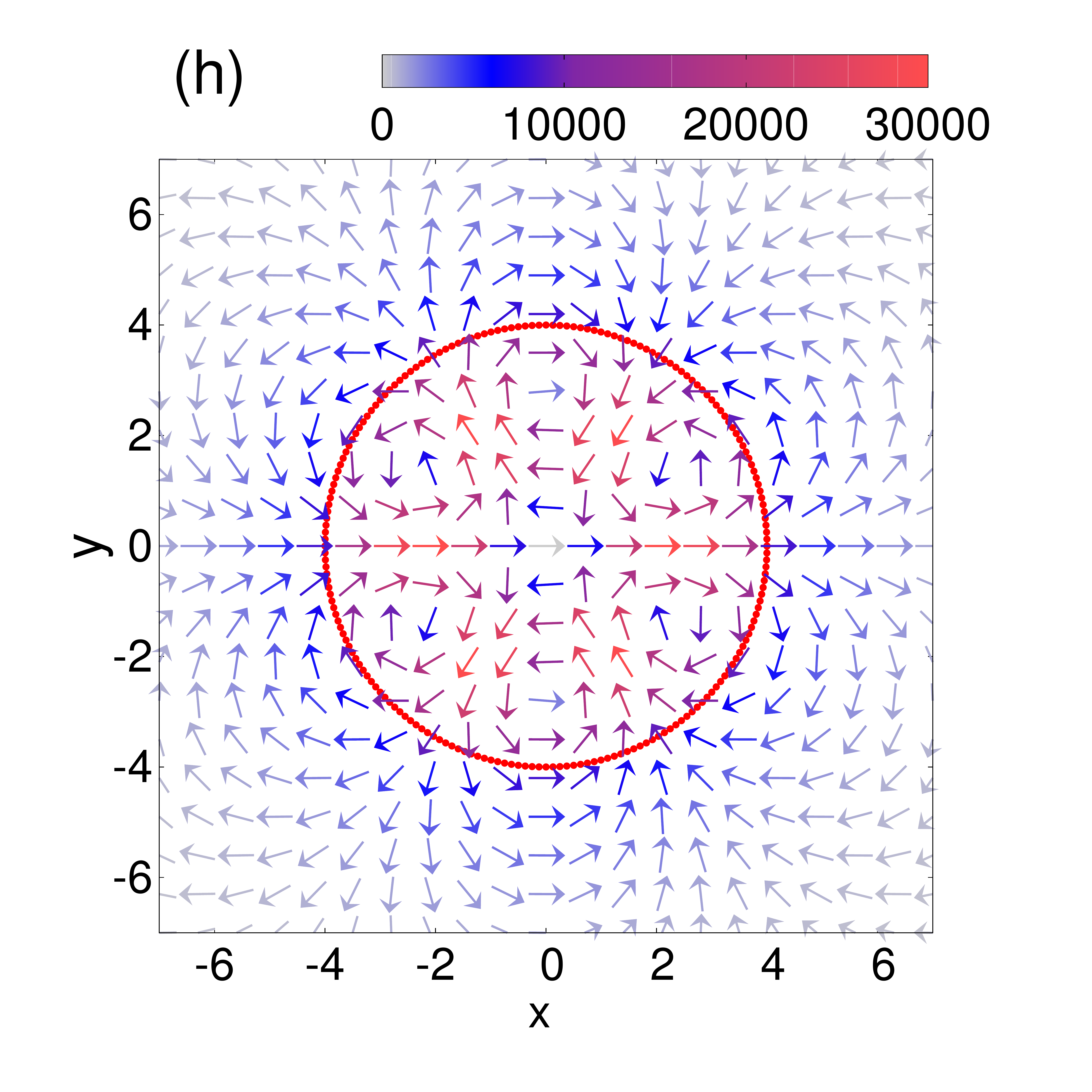}}
\end{minipage}

\begin{minipage}{0.32\linewidth}
{\includegraphics[width=1\linewidth]{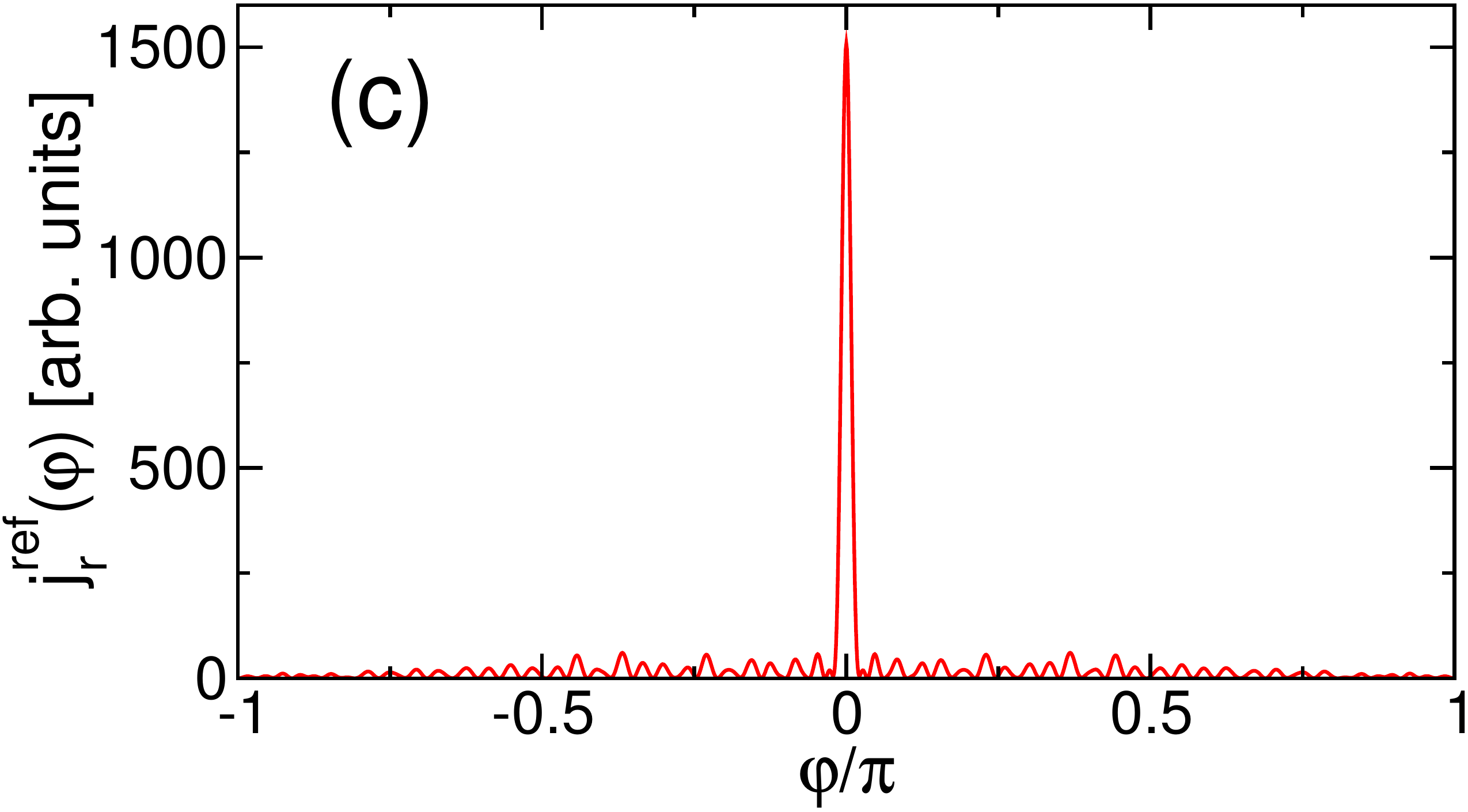}}
\end{minipage}
\begin{minipage}{0.32\linewidth}
{\includegraphics[width=1\linewidth]{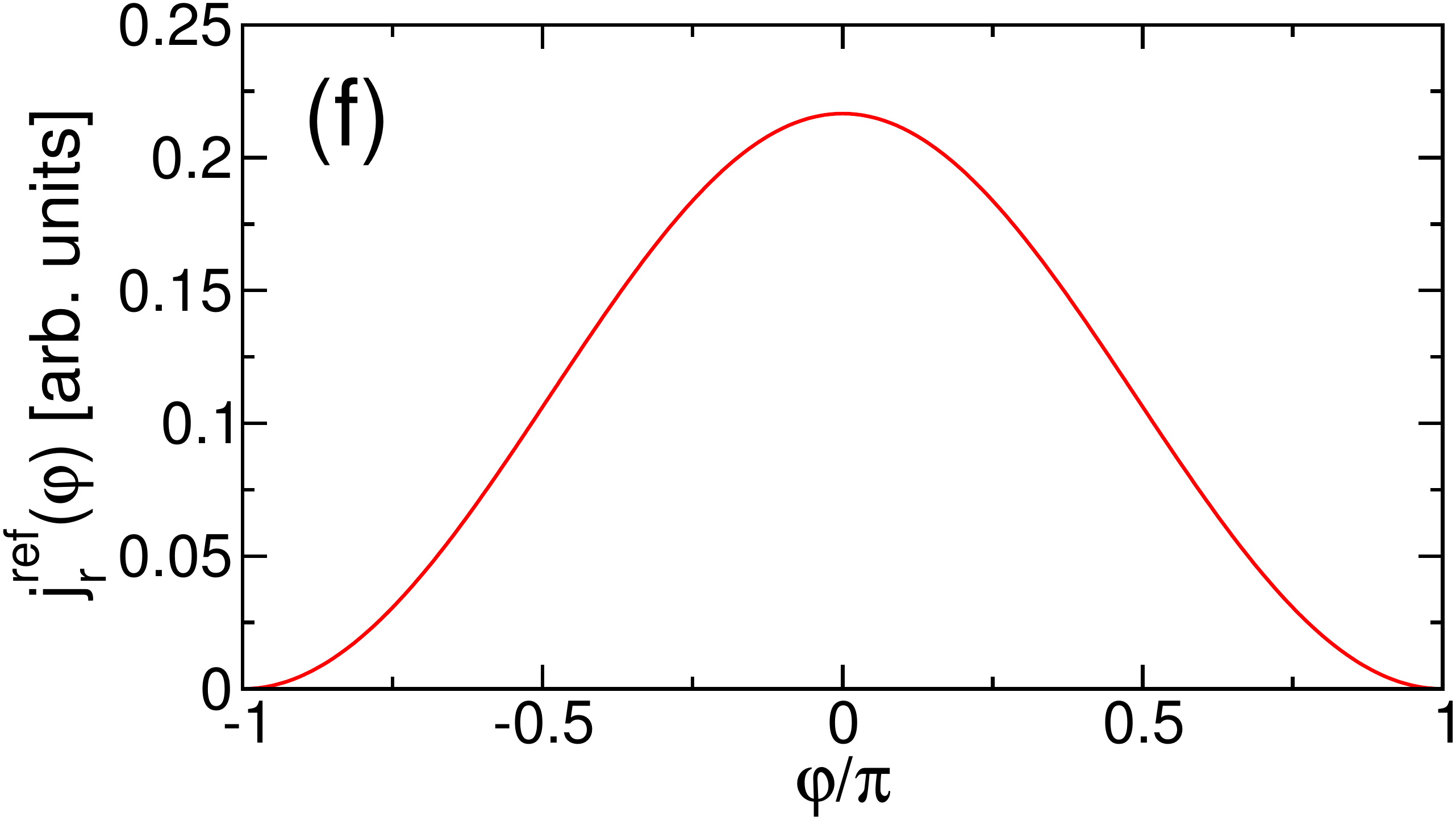}}
\end{minipage}
\begin{minipage}{0.32\linewidth}
{\includegraphics[width=1\linewidth]{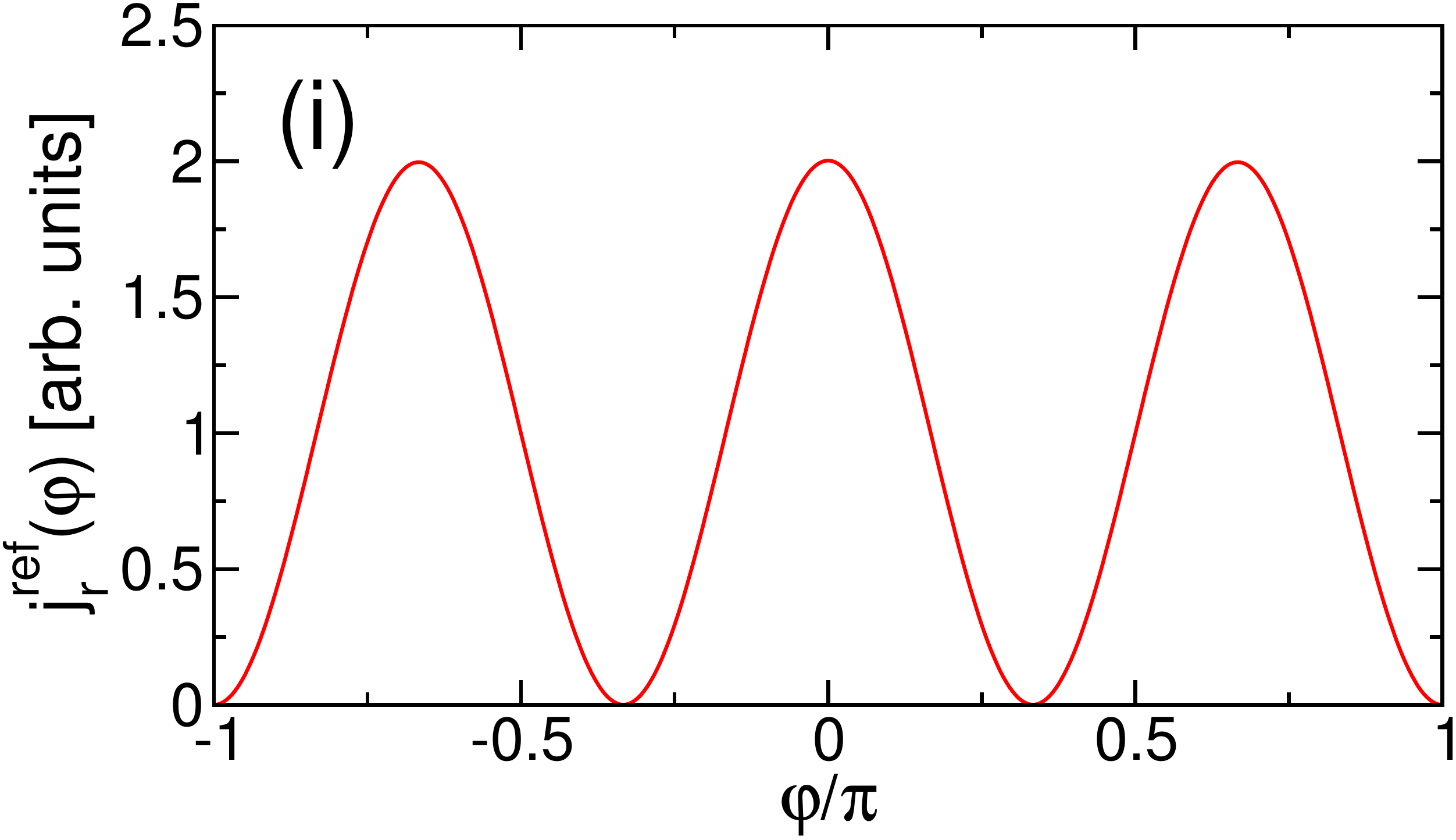}}
\end{minipage}
\caption{(color online) Lensing and particle confinement by the gated region.
 Density \(n=\psi^\dagger \psi\) (top row), current density \(\mathbf{j}=\psi^\dagger \mathbf{\sigma}\psi\) (middle row) and far field radial component of the reflected  current \(j_r^\mathrm{ref}\) (bottom row) for (a)-(c) \(R=100\), \(E=0.5\), \(V=1\), (d)-(f) \(R=2\), \(E=0.1\), \(V=1\), and (g)-(i) \(R=4\), \(E=0.02813\), \(V=1\). (a) For large \(R\) and moderate \(E\) refraction increases \(n\) along two caustics. (b) The electron flow around the dot causes (c) peaked forward scattering.  The \(a_0\) mode has (d) maximum electron density in the center (e) two vortices in the current field and, (f) preferred forward scattering. The \(a_1\) mode has (g) a ring shaped electron density, (h) six vortices in the current field and, (i) three preferred scattering directions. }
\label{figure2}
\end{figure*}
 
\begin{figure*}
\begin{minipage}{0.49\linewidth}

\includegraphics[width=0.55\linewidth]{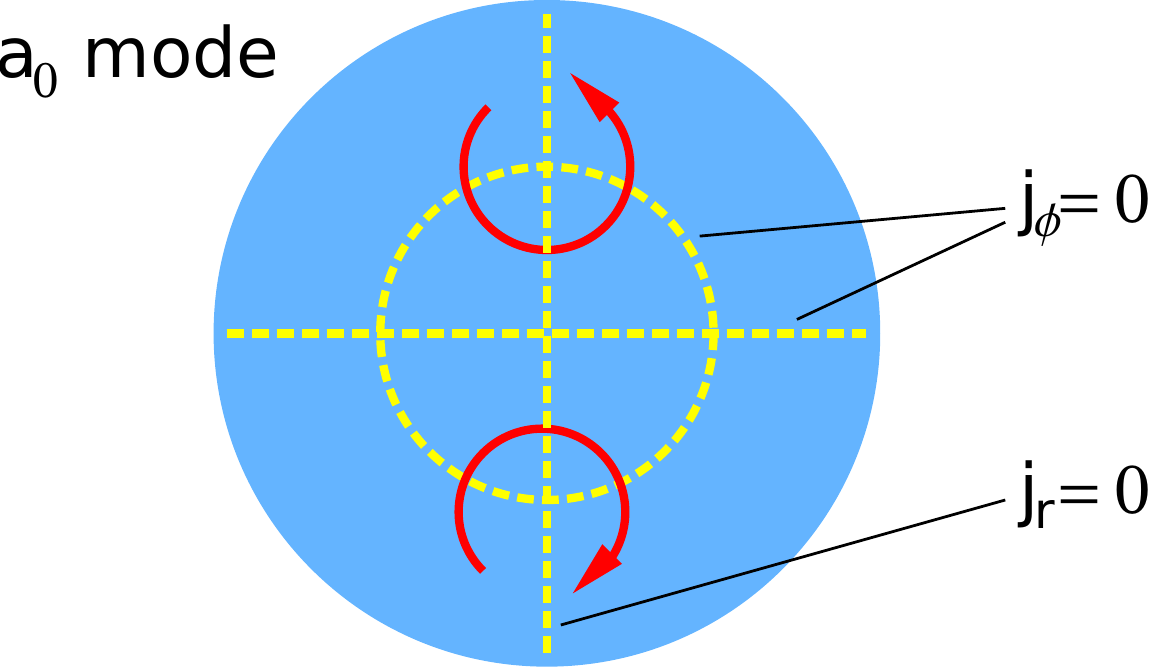}
\end{minipage}
\begin{minipage}{0.49\linewidth}
\caption{(color online) Schematic representation of the current in the gated region for the \(a_0\) mode. The angular dependence  of the current (see Eq. (\ref{modecurr})) leads to \(j_r=0\)   on the \(y\)-axis and to  \(j_\phi=0\)   on the \(x\)-axis. The radial dependence in Eq. (\ref{modecurr}) leads to a concentric circle where \(j_\phi=0\). Around the intersection of this circle with the y-axis two vortices are formed which stream the electron through the middle of the gated region.}
\label{figure3}
\end{minipage}

\end{figure*} 
 
\begin{figure*}
\includegraphics[width=0.95\linewidth]{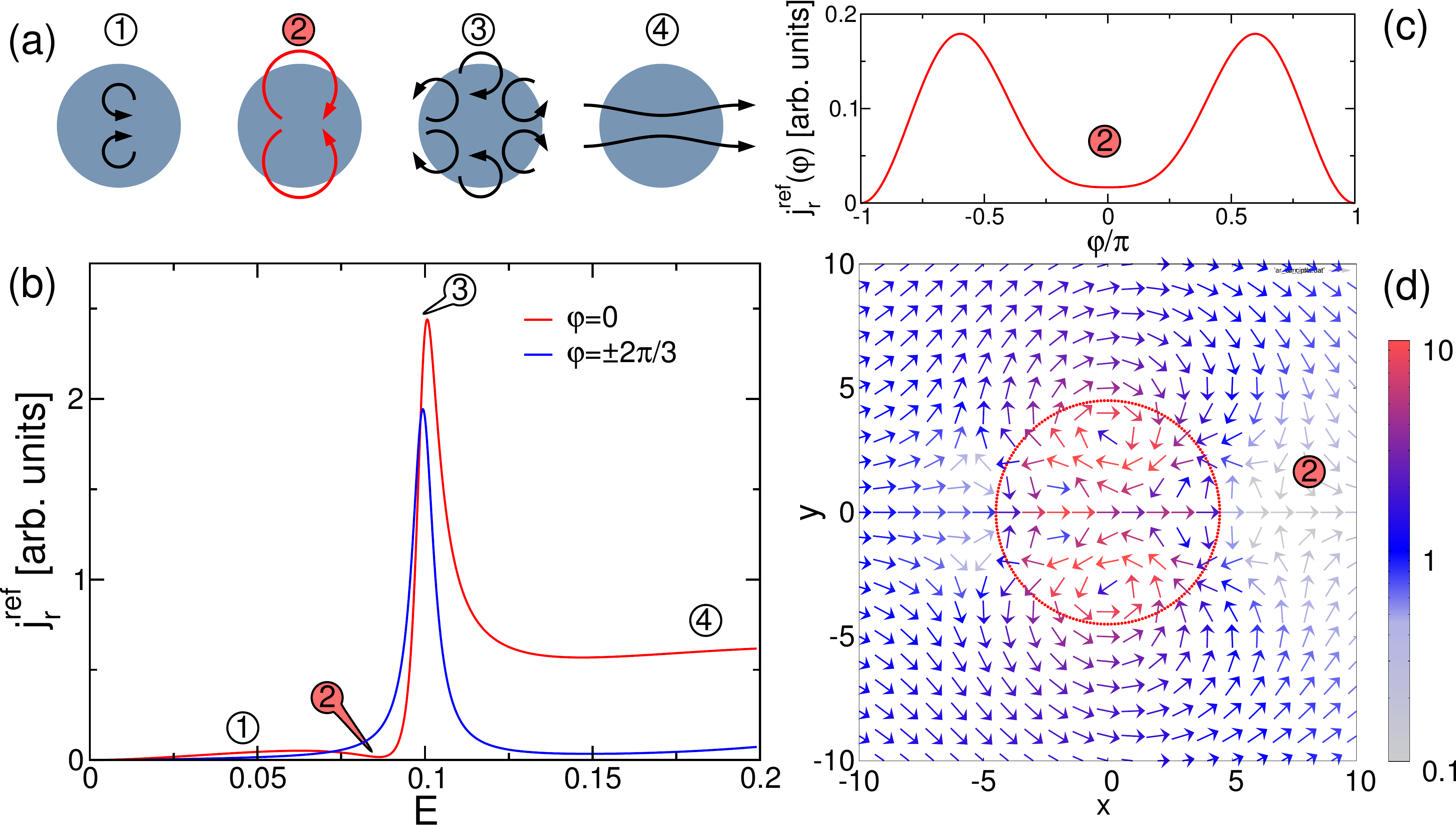}

\caption{(color online) Depletion of Klein tunneling by Fano resonance. (a) Schematic representation of the vortices in the current field: (1) and (3) Vortex pattern of the \(a_0\) and \(a_1\) mode. (2) Enhanced counterrevolving top and bottom vortices of the \(a_1\) mode (red) deplete Klein tunneling. (4) Top and bottom vortices are suppressed and electrons stream through the gated region. (b) Far field radial reflected current \(j_r^\mathrm{ref}\) for \(\varphi=0\) (forward scattering) and \(\varphi=\pm 2 \pi /3\)  as a function of the energy \(E\) for a gated region with bias \(V=1\) and radius \(R=4.5\). For \(\varphi=0\) the lineshape  is asymmetric due to interference between the \(a_0\) and \(a_1\) mode while for \(\varphi=\pm 2 \pi /3\) it is Lorentzian.   (c) \(j_r^\mathrm{ref}\) in the far field  as a function of the scattering angle \(\varphi\) and (d) current density \(\mathbf{j}=\psi^\dagger \mathbf{\sigma}\psi\) in the near field for \(E=0.0863\), \(R=4.5\) and \(V=1\). Marker 2 indicates the suppression of Klein tunneling.}
\label{figure4}

\end{figure*}

Let us now analyze the scattering by a gated region in graphene in more detail. For large radius compared to the wavelength of the electron scattering shows features known from ray optics. 
Refraction inside the gated region gives rise to two caustics which coalesce in a cusp (Fig.~\ref{figure2}a). The caustics have already been studied in Ref. \onlinecite{CPP07}, including a ray-optical derivation for them.
Obviously, the boundary of the gated region acts as a lens focusing the electron beam. The radiation characteristic, given by the radial component of the far field reflected current \(j_r^\mathrm{ref}\), shows the absence of backscattering and a peaked forward scattering (Fig.~\ref{figure2}c). This is surprising as the incident electron impacts non-normally on parts of the circular gated region which should give rise to reflection. It turns out that in the near field the electron flows around the dot (Fig. 2b), which results in the peaked forward scattering.  

For small radius and low energy of the incident electron scattering resonances appear at specific values of \(E\) and \(R\) where one of the \(a_m^\mathrm{r}\) reaches unity. In between the resonances only the \(a_0\) mode is excited off-resonantly. For small \(\rho\) the coefficient \(a_m^\mathrm{r}\) gives rise to a series of resonances for \(\rho N =j_{m,s}\), where \(j_{m,s}\) is the \(s^{th}\) zero of  \(J_m\). For small \(\rho\) (or \(E\)) and higher orders \(m\) their linewidth shrinks (hence, the resonances \(a_{m\geq 1}\) are not resolved for small \(E\) in Fig. 1b,d). In the limit \(E=0\) the resonances are located at \(RV=j_{m,s}\) with zero linewidths, in agreement with the  bound states found for a dot in unbiased graphene.\cite{BTB09,TOG10} If only one mode is excited the electron density,
\begin{equation}
n=\left|a_m^\mathrm{t}\right|^2 \left( J_m^2(qr)+J_{m+1}^2(qr)\right),
\end{equation}
 is radially symmetric (Fig.~\ref{figure2}d,g).  For resonant scattering the electron density increases dramatically inside the dot (Fig.~\ref{figure2}g; note the different intensity scale). This indicates temporary particle confinement in the gated region. The current field inside the dot is
\begin{align}
\mathbf{j}=&\left|a_m^\mathrm{t} \right|^2 \alpha \left( J_{m+1}^2(qr)+J_m^2(qr) \right) \cos ((2m+1)\phi) \mathbf{\hat{e}}_r \nonumber \\
+& \left|a_m^\mathrm{t} \right|^2\alpha \left(J_{m+1}^2(qr)-J_m^2(qr)\right) \sin ((2m+1)\phi) \mathbf{\hat{e}}_\phi . \label{modecurr}
\end{align} 
An analysis of this expression (see Fig.~\ref{figure3} for the \(a_0\) mode) reveals a vortex pattern symmetric to the \(x\)-axis which dominates the near-field (Fig.~\ref{figure2}e,h). The incident wave is fed into vortices, which trap the electron. Note that the particle is confined in vortices, not by total internal reflection.\cite{BTB09} The vortex pattern of the mode \(a_m\)  is dominated by \(2(2m+1)\) vortices close to the boundary of the circle which are reflected by  \(2m+1\) preferred scattering directions in the far field. The radial reflected current 
\begin{equation}
j_r^\mathrm{ref} \sim \cos ((2m+1)\varphi) +1
\end{equation}
shows that for the \(a_0\) mode only forward scattering is favoured (Fig.~\ref{figure2}f), while for higher modes more preferred scattering directions emerge (Fig.~\ref{figure2}i for \(a_1\)). 

For small electron energies the \(a_0\) mode is relatively broad compared to the sharp resonances of higher modes. Constructive and destructive interference between a resonant \(a_m\) mode and the off-resonant \(a_0\) mode can give rise to Fano resonances similar to the  effect of the hybridization between a continuum and a discrete level in electronic transitions.\cite{Fano61} In optics they are routinely used to tailor properties of plasmonic structures\cite{HSV10,LZA10} and have recently been identified for Mie scattering by a sphere,\cite{TFM08,TMK12} where they allow to switch from forward to backward scattering. The scattering efficiency reveals no Fano signatures but the  angle dependent scattering which depends also on the phases of the scattering coefficients encompasses interference effects. Remarkably we find that over a Fano resonance a small variation of parameters can lead to dramatic changes in the near field, such as an inversion of the vortex pattern which causes in the far field a change in the preferred scattering direction and even leads to a suppression of Klein tunneling and electron depletion behind the quantum dot. Figure~\ref{figure4} gives the example of the \(a_1\) resonance (preferred scattering angles \(\varphi=0,\pm 2\pi/3\)), over an off-resonant \(a_0\) background (preferred scattering angle \(\varphi=0\)). For \(\varphi=0\) the contributions of the \(a_0\) and \(a_1\) mode can have the same  amplitude and interference causes an asymmetric lineshape of \(j_r^\mathrm{ref}\) across the resonance with enhancement above and suppression below. This can be understood from  the vortex structure in the near field. Above the resonance destructive interference suppresses counterrevolving vortices and electrons stream through the center of the gated region. Below the resonance the couterrevolving vortices are enhanced and suppress forward scattering (Fig.~\ref{figure4}a,b). Only side scattering remains in the far field (Fig.~\ref{figure4}c) and, most notably, in the near field the current field shows a depletion of Klein tunneling at the dot (Fig.~\ref{figure4}d).

\section{conclusions}

To sum up, we demonstrated that a circular gated region in graphene may act as a lens focusing a Dirac electron beam. Scattering resonances enable particle confinement and interference effects may switch forward scattering on and off. Our results, which should be verified by conduction measurements or by mapping the electron density at the gated region exposed to a plane electron wave by a scanning-tunneling microscope, suggest a strong analogy to light scattering by a sphere. The gated region may act as a switch in graphene, which by tuning its potential manipulates the preferred scattering direction. This opens the possibility for the design of graphene-based circuitry by spatially structured electric biasing.

\section*{acknowledgement}

This work was supported by the DFG Priority Programme 1459 Graphene and SFB-TR 24. We thank C. Schulz for discussions.


\end{document}